\documentclass[superscriptaddress,twocolumn,showpacs]{revtex4}
\usepackage[dvipdfmx]{graphicx}
\usepackage{color}
\usepackage{dcolumn}
\usepackage{amsmath}
\usepackage{dsfont}
\usepackage{epsfig,color}
\usepackage{soul}
\usepackage{amssymb}

\newcommand{\be}{\begin{equation}}
\newcommand{\ee}{\end{equation}}

\begin{document}

\title{Scattering of Waves by Impurities in Precompressed Granular Chains}

\author{Alejandro J. Mart\'inez}

\affiliation{Oxford Centre for Industrial and Applied Mathematics,
Mathematical Institute, University of Oxford, Oxford OX2 6GG, United Kingdom}

\author{Hiromi Yasuda}
\thanks{The first two authors contributed equally.}
\affiliation{Aeronautics and Astronautics, University of Washington, Seattle, WA 98195-2400, USA}

\author{Eunho Kim}

\affiliation{Aeronautics and Astronautics, University of Washington, Seattle, WA 98195-2400, USA}

\author{P. G. Kevrekidis}

\affiliation{Center for Nonlinear Studies and Theoretical Division, Los Alamos
National Laboratory, Los Alamos, NM 87544, USA}

\affiliation{Department of Mathematics and Statistics, University of
Massachusetts, Amherst, Massachusetts 01003-4515, USA}

\author{Mason A. Porter \footnote{Corresponding author}}

\affiliation{Oxford Centre for Industrial and Applied Mathematics,
Mathematical Institute, University of Oxford, Oxford OX2 6GG, United Kingdom}

\affiliation{CABDyN Complexity Centre, University of Oxford, Oxford OX1 1HP, United Kingdom}

\author{Jinkyu Yang}

\affiliation{Aeronautics and Astronautics, University of Washington, Seattle, WA 98195-2400, USA}




\begin{abstract}

We study scattering of waves by impurities in strongly precompressed granular chains. We explore the linear scattering of plane waves and identify a closed-form expression for the reflection and transmission coefficients for the scattering of 
the waves from both a single impurity and a double impurity. For single-impurity chains, we show that, within the transmission band of the host granular chain, high-frequency waves are strongly attenuated (such that the transmission coefficient vanishes as the wavenumber $k\rightarrow \pm\pi$), whereas low-frequency waves are well-transmitted through the impurity. For double-impurity chains, we identify a resonance --- enabling full transmission at a particular frequency --- in a manner that is analogous to the Ramsauer--Townsend (RT) resonance from quantum physics.  We also demonstrate that one can tune the frequency of the RT resonance to any value in the pass band of the host chain. We corroborate our theoretical predictions both numerically and experimentally, and we directly observe complete transmission for frequencies close to the RT resonance frequency. 
Finally, we show how this RT resonance can lead to the existence of reflectionless modes even in granular chains (including disordered ones) with multiple double impurities.

\end{abstract}

\maketitle



\section{Introduction}

One-dimensional (1D) granular crystals (i.e., granular chains) consist of closely packed chains of elastically colliding particles. This setup has been used as a testbed for the investigation of wave phenomena in chains of strongly nonlinear oscillators, and the interplay between nonlinearity and discreteness in granular chains has inspired the exploration of a diverse set of coherent structures, including traveling waves, breathers, and dispersive shock waves~\cite{Nesterenko:book,Sen:PR2008,pgk:2011}.  Granular crystals can be constructed from a wide variety materials of different types and sizes, so their properties are very tunable, and they thus provide a versatile type of metamaterial for both fundamental physical phenomena and applications~\cite{Nesterenko:book,Sen:PR2008,Daraio:PRE2006,Coste:PRE1997}.  

Granular crystals have been used for the investigation of numerous structural and material heterogeneities on nonlinear wave dynamics. This includes the role of defects \cite{PhysRevE.57.2386,Hascoet:EPJB2000,Hong:APL2002,Theocharis:PRE2009} (including in experimental settings~\cite{Job:PRE2009,Many}); the scattering from interfaces between 
two different types of particles \cite{Nesterenko:PRL2005,Daraio:PRL2006,PhysRevE.82.061303}; and wave propagation
in decorated and/or tapered chains \cite{Doney:PRL2006,Harbola:PRE2009}, chains of diatomic and triatomic units \cite{Porter:PRE2008,Porter:PhysicaD2009,Herbold:Mechanica2009,Molinari:PRE2009,staros1,staros2,staros3,boechler4}, and quasiperiodic and random configurations \cite{Sokolow:AnnPhys2007,Chen:PhysicaB2007,Fraternali:MAMS2010,Ponson:PRE2010,martinez2014,theocharis2015,yous2015}; and much more. One can model strongly compressed granular chains as a type of Fermi--Pasta--Ulam (FPU) lattice, and granular chains have been employed in studies of phenomena such as equipartition (see, e.g., \cite{szel2013,zhang2015}).

Granular chains also provide prototypes for numerous potential engineering applications~\cite{PT2015}. A few examples include shock and energy absorbing layers \cite{Daraio:PRL2006,Hong:PRL2005,Fraternali:MAMS2010}, sound-focusing devices and delay lines \cite{Spadoni:PNAS2010}, actuators \cite{Khatri:SPIE2008}, vibration absorption layers \cite{Herbold:Mechanica2009}, sound scramblers \cite{Daraio:PRE2005,Nesterenko:PRL2005}, and acoustic switches and logic gates~\cite{daraio_natcom}.  

The study of disordered granular crystals is also becoming increasingly popular. Important themes in such studies have been transport properties of wavepackets and solitary waves and the interplay between disorder (especially in the context of
Anderson localization), discreteness, and nonlinearity~\cite{Ponson:PRE2010,martinez2014,theocharis2015,yous2015}. These themes are also relevant to a wide variety of other nonlinear lattice models~\cite{Flach:ArxivRep2014-1,Flach:ArxivRep2014-2}.

To get a handle on disordered granular chains, it is useful to start with a simpler setting in which one or a few defects occur
within an otherwise homogeneous (``host'') lattice~\cite{Theocharis:PRE2009}.
In this context, scattering due to inhomogenities is a fundamental consideration when studying wave propagation in complex media~\cite{scatteringbook,Miroshnichenko:RMP2010}. This is especially important when the scales of the waves and 
those of the inhomogeneities (i.e., impurities or defects) are comparable, as interactions in such situations can lead to very rich dynamics. Pertinent phenomena include the formation of localized modes~\cite{Flach:PR2008,DSopt}, Fano resonances~\cite{Fano1,Fano2}, cloaking~\cite{cloaking1,cloaking2}, and many other examples of broad interest 
across numerous branches of physics.

In the present paper, we use theory, numerical computations, and experiments in the linearized and weakly nonlinear regimes to explore the scattering of a plane wave from a single impurity and a double impurity in a granular chain. A key finding is that an analog of the well-known Ramsauer--Townsend (RT) effect can occur in 
granular chains. An RT resonance is a prototypical mechanism
that enables scattering transparency in quantum 
mechanics~\cite{Sakurai}. In its most recognizable form, it consists of the presence of a sharp minimum in 
the electron scattering cross-section at low energies for scattering with rare gases. The RT effect has been observed experimentally in many scenarios involving quantum tunneling, including $e^-$--Ar scattering~\cite{RTEffect1} and positron--Ar scattering~\cite{RTEffect2}, $e^-$--methane scattering~\cite{RTEffect3}, and others. 
When used in mechanical systems, the implication of the RT effect is equally 
significant. One possible application is embedding foreign objects, such as sensors, in systems so that they induce minimal interference with the existing structures. 
This has the potential to be very useful for applications in structural health monitoring.  

The remainder of our paper is organized as follows. In Sec.~\ref{Sec2}, we introduce the fundamental equations that govern the dynamics of driven granular crystals. In Sec.~\ref{scat}, we solve, in closed form, the linear scattering problems of a
single impurity and a double impurity embedded in a homogeneous (``host'') granular chain. 
For double impurities, we demonstrate that an effect analogous to an RT resonance occurs in a well-defined region of parameter space. We use both numerical simulations and laboratory experiments to corroborate our theoretical results. In Sec.~\ref{sec:comparison}, we discuss and compare the results from our theory, computations, and experiments. In Sec.~\ref{sec:multiple}, we use numerical simulations to explore disordered granular chains, which include a large number of impurities. We demonstrate numerically that strongly precompressed chains with multiple impurities can admit solutions that consist of reflectionless modes (i.e., generalizations of the RT resonances). Finally, in Sec~\ref{conclusion}, we conclude and offer some directions for future work.


\section{Driven Granular Crystals}\label{Sec2}

One can describe a 1D crystal of $2N+1$ spherical particles as a chain of nonlinearly coupled oscillators with Hertzian interactions between each pair of particles \cite{Nesterenko:book,Sen:PR2008,pgk:2011}. The system is thus modeled using the following equations of motion:
\begin{equation}
	\ddot{u}_n = \frac{A_n}{m_n}[\Delta_n+u_{n-1}-u_n]_+^{3/2}-
\frac{A_{n+1}}{m_n}[\Delta_{n+1}+u_n-u_{n+1}]_+^{3/2}\,, 
\label{Hertz}
\end{equation}
where $m_n$ is the mass of the $n$th particle, $u_n$ is the displacement of the $n$th particle
(where $n\in \{-N,-N+1,{\ldots},N\}$) measured from its
equilibrium position, the pairwise interaction parameter $A_n$ depends on 
the geometry and elasticity of particles in the $n$th and $(n-1)$th 
positions~\cite{Nesterenko:book},
\begin{equation}
	\Delta_n = \left(\frac{F_0}{A_n}\right)^{2/3}
\end{equation}
is the change in displacement between centers of neighboring particles due to the static load $F_0$, and the bracket $[\cdot]_+$ is defined as 
\begin{equation}
	[x]_+ = \left\{
		\begin{array}{lcc}
	x\,, & \text{if} & x>0\\
	0\,, & \text{if} & x\leq0
		\end{array}\right.\,.
\end{equation} 

We consider a chain that is compressed initially by two plates placed at the boundaries. This yields the following boundary conditions: 
\begin{align}
	u_{-(N+1)} &= \psi_l(t)\,,\\
	u_{N+1} &= \psi_r(t)\,.
\end{align}
We focus on a situation in which the chain is driven periodically from one side and the other side is at rest. That is, $\psi_r(t) = 0$ and $\psi_l(t) = d \sin(2\pi f t)$, where $d$ and $f$, respectively, are the amplitude and frequency of the external driving. 

We are interested in chains that are homogeneous except for a few particles (i.e., impurities) in the bulk.  We consider two cases: (i) a single impurity and (ii) a double impurity (in which the impurities are adjacent particles). The interaction parameter $A_n$ can take one of four possible values (depending on the type of spheres that are in contact). These values are
\begin{equation}
	A_n = \left\{
		\begin{array}{lc}
	A_{11}\equiv \frac{E_{1}\left(2r_{1}
	\right)^{1/2}}{3(1-\nu_{1}^2)}\,,
	& \text{(type-1, type-1)}\\
	A_{12}\equiv \frac{4E_{1}E_{2}\left(\frac{r_{1}r_2}{r_{1}+r_2}
	\right)^{1/2}}{3\left[E_{1}(1-\nu_{2}^2)+E_{2}(1-\nu_{1}^2)\right]}\,,
	& \text{(type-1, type-2)}\\
	A_{22}\equiv \frac{E_{2}\left(2r_{2}
	\right)^{1/2}}{3(1-\nu_{2}^2)}\,,
	& \text{(type-2, type-2)}\\
	\frac{2E_1 r_{1}^{1/2}}{3(1-\nu_1^2)}\,,
	& \text{(type-1, wall)}
		\end{array}\right.\,,
\end{equation}
where $E_{1,2}$, $\nu_{1,2}$, and $r_{1,2}$ are, respectively, the elastic modulus, the Poisson
ratio, and the radii of the type-1 and type-2 particles. The particle masses are $m_1$ and $m_2$. We assume that the mechanical properties of the elastic plates at the boundaries are the same as for type-1 particles. 
The radius of an impurity particle is $r_2 = \alpha r_1$, where $\alpha > 0$ is the ratio between the radii of the two types of spheres. If we assume that type-1 and type-2 particles have identical densities 
(i.e., $\rho_1=\rho_2$), then $\alpha <1$ implies that the impurities are lighter than the particles in the host homogeneous chain, whereas $\alpha>1$ implies that the impurities are heavier.


\section{Scattering between linear waves and impurities}\label{scat}

Depending on the relative magnitudes of $\Delta_n$ and $|u_n-u_{n+1}|$, the effective nonlinearity in Eq.~\eqref{Hertz} can be either strong or weak. In particular, for sufficiently strong static precompression or sufficiently small-amplitude vibrations in the crystal, $\Delta_n\gg |u_{n-1}-u_n|$, so the nonlinearity is very weak. If one ignores the nonlinearity entirely, we have a harmonic interaction between the particles, so the dynamics can be described by the equation
\begin{equation}
	m_n\ddot{u}_n = B_{n}u_{n-1}+B_{n+1}u_{n+1}-(B_n+B_{n+1})u_n\,,
\label{Linear}
\end{equation}
which corresponds to Eq.~\eqref{Hertz} linearized about the equilibrium state. Consequently,
\begin{equation}\label{equ9}
	B_{n} =\frac{3}{2}A_n\Delta_n^{1/2} =  \frac{3}{2}A_n^{2/3}F_0^{1/3}\,.
\end{equation}
One can express solutions of Eq.~\eqref{Linear} in terms of a complete set of eigenfunctions of the form $u_n = v_n e^{i\omega t}$, where $\omega$ is the eigenfrequency. It is well-known that without impurities --- i.e., for a completely homogeneous crystal with $m_n=m$ and $B_n=B$ --- that $v_n = e^{ikn}$, so there is a single acoustic branch of solutions with eigenfrequency 
\begin{equation}
	\omega =\sqrt{\frac{2B}{m}\left[1-\cos(k)\right]} \in [0,\Omega]\,,
\end{equation}	
where $m$ is the mass, $k$ is the wavenumber, and $\Omega = \sqrt{\frac{4B}{m}}$. When impurities are introduced into a host chain, localized or resonant linear modes can arise (depending on the characteristics of the impurities \cite{Theocharis:PRE2009}.) For light impurities (i.e., $\alpha < 1$), one expects localized modes whose frequencies are larger than the upper bound $\Omega$ of the linear spectrum. For heavy impurities (i.e., $\alpha > 1$), by contrast, one expects impurity modes with frequencies in the linear spectrum, and one thus expects resonant modes
with extended linear eigenmodes.

\begin{figure}[t]
\includegraphics[height=4.5cm]{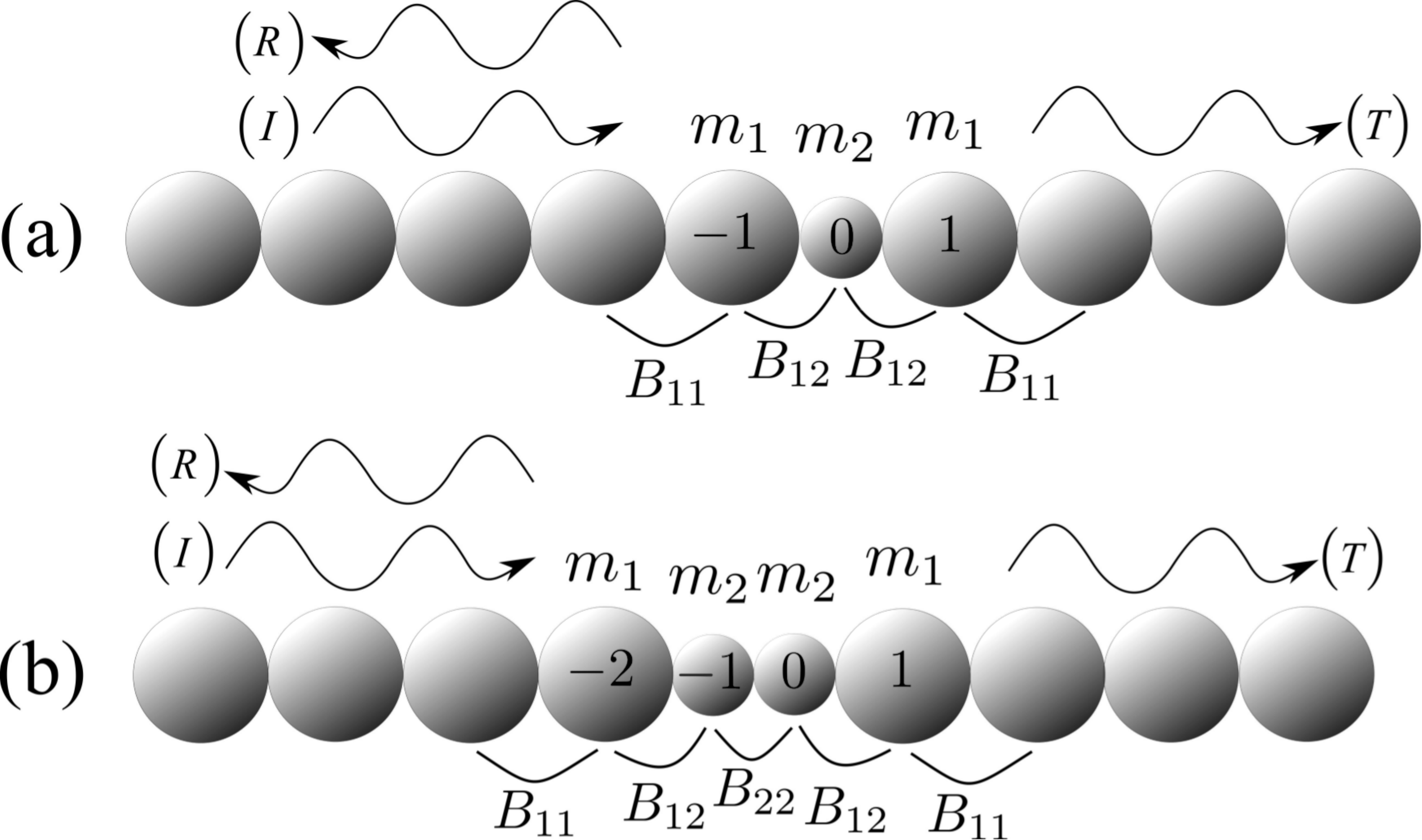}
\caption{Schematic of a homogeneous granular chain with (a) one impurity and (b) two
contiguous impurities (i.e., a double impurity). The incident wave is 
characterized by $I$, the reflected wave by $R$, and the transmitted wave by $T$. We label the identities of the particles with integers. We calculate the parameters $B_{ij} = \frac{3}{2}A_{ij}^{2/3}F_0^{1/3}$ from the static precompression and the interactions between consecutive particles. 
}
\label{f4-scattering}
\end{figure}

\subsection{Theory} \label{sec:theory}

We are interested in studying scattering processes between a plane wave
$e^{i(kn-\omega t)}$ and both single impurities and double impurities in the linear
regime. In Fig.~\ref{f4-scattering}, we show schematics for chains with single and double impurities.
We treat an impurity particle as a perturbation of a host particle: an impurity particle has radius $r_2 = \alpha r_1$, where $r_1$ is the radius of a host particle in the chain. We focus on $\alpha\in(0,2]$.
The value of the parameter $\alpha$ determines the mass of an impurity and the 
values of the interaction coefficients $A_{n}$ between neighboring particles. For double impurities, we only consider the ``symmetric'' case in which both impurities are the same type of particle (and hence have the same radius).

To solve the scattering problem in the linear regime, it is convenient to use complex quantities rather than real ones. We write~\cite{Miroshnichenko:RMP2010}
\begin{equation}
	u_n = \left\{
		\begin{array}{lcr}
		e^{i(kn-\omega t)}+Re^{-i(kn+\omega t)}\,,&\quad\text{if}
		&n\leq0\\
		Te^{i(kn-\omega t)}\,,&\quad\text{if}&n>0
		\end{array}
	\right.\,,
\label{ansatz}
\end{equation}
which represents an incident plane wave producing reflected and transmitted waves due to the interaction with the impurity. 
We thereby define a transmission coefficient $|T|^2$ and a reflection coefficient $|R|^2$. Note that $|R|^2 + |T|^2$ need not equal $1$ because both $|T|^2$ and $|R|^2$ are based on the norm of the displacement, which is not a 
conserved quantity of either Eq.~\eqref{Hertz} or Eq.~\eqref{Linear}. Intuitively, $|T|^2$ and $|R|^2$ are still ``complementary'' quantities,  as a decrease in one is accompanied by an increase 
in the other (and vice versa). To have $|R|^2 + |T|^2 = 1$ for all parameter values, one would need to instead define $|R|^2$ and $|T|^2$ in terms of the energy density. The total energy is conserved by the dynamics, though it is much harder to measure experimentally than other quantities (e.g., velocity). 
Given Eq.~\eqref{ansatz}, the velocity is $\dot{u}_n = -\omega u_n$. Therefore, if we defined $|T|^2$ and $|R|^2$ in terms of $\dot{u}_n$ rather than $u_n$, we would obtain the same results because $\dot{u}_n$ and $u_n$ differ only by the constant factor $-\omega$. We therefore define reflection and transmission coefficients in terms of displacement, which allows us to compare analytical results directly with not only computations but also laboratory experiments, for which we compute the coefficients in terms of velocity (see Secs.~\ref{sec:simulation} and~\ref{sec:experiment}).

We substitute the ansatz \eqref{ansatz} into Eq.~\eqref{Linear} near the impurities and do a straightforward calculation to obtain the following linear system of equations for $T$ and $R$:
\begin{align}
\left(\begin{array}{cc}
       \beta_{(i),(ii)}&\delta_{(i),(ii)}\\
       \eta_{(i),(ii)}&\gamma_{(i),(ii)}
       \end{array}
\right)\left(\begin{array}{c}
       T_{(i),(ii)}\\
       R_{(i),(ii)}
       \end{array}
\right)
=\left(\begin{array}{c}
       \epsilon_{(i),(ii)}\\
       \zeta_{(i),(ii)}
       \end{array}
\right)\,,
\label{matrixeq}
\end{align}
where the subscripts $(i)$ and $(ii)$, respectively, indicate chains with single and double impurities. 

For a single-impurity chain, the parameters in Eq.~\eqref{matrixeq} are
\begin{align*}
	\beta_{(i)} &=\bar{\Omega}+B_{12}(2-e^{ik})\,,\\
	\gamma_{(i)} &=-B_{12}e^{ik}\,,\\
	\delta_{(i)} &=B_{12}e^{-ik} \,,\\
	\eta_{(i)} &= B_{12}\,,\\
	\epsilon_{(i)} &= B_{11}e^{2ik}-(B_{11}+B_{12}+\bar{\Omega})e^{ik}\,,\\
	\zeta_{(i)} &= -B_{11}e^{-2ik}+(B_{11}+B_{12}+\bar{\Omega})e^{-ik}\,,
\end{align*}
where $\bar{\Omega}=-\frac{2B_{11}m_2}{m_1}[1-\cos(k)]$. Solving Eq.~\eqref{matrixeq} yields the reflection and transmission coefficients:
\begin{widetext} 
\begin{align}
	|R_{(i)}|^2 &= \left|\frac{B_{11}(B_{11}-B_{12})m_2 -
	(2B_{11}-B_{12})(B_{11}m_2-B_{12}m_1)e^{ik}+B_{11}(B_{11}m_2-B_{12}m_1)e^{2ik}}
	{B_{11}^2 m_2
	e^{ik}+(B_{11}-B_{12})(B_{11}m_2-B_{12}m_1)e^{3ik}-B_{11}(2B_{11}m_2-2B_{12}m_1-
	B_{12}m_2)e^{2ik}}\right|^2\,,\nonumber\\
	|T_{(i)}|^2 &= \left| \frac{B_{11}B_{12}m_1(1+e^{ik})}
	{B_{11}^2
	m_2+(B_{11}-B_{12})(B_{11}m_2-B_{12}m_1)e^{2ik}-B_{11}(2B_{11}m_2-2B_{12}m_1-
	B_{12}m_2)e^{ik}}\right|^2\,.
\label{single}
\end{align}
\end{widetext}

For a double-impurity chain, we follow the same procedure and use the parameters
\begin{align*}
	\beta_{(ii)} &=\bar{\Omega}+B_{12}(1-e^{ik})+B_{22}\,,\\
	\gamma_{(ii)} &=-B_{22}e^{ik}\,,\\
	\delta_{(ii)} &=B_{22}e^{-ik}\,,
\end{align*}
\begin{align*}
	\eta_{(ii)} &= B_{22}\,,\\
	\epsilon_{(ii)} &= B_{12}e^{2ik}-(B_{12}+B_{22}+\bar{\Omega})e^{ik}\,,\\
	\zeta_{(ii)} &= -B_{12}e^{-2ik}+(B_{12}+B_{22}+\bar{\Omega})e^{-ik}
\end{align*}
in Eq.~\eqref{matrixeq}. Note that $\bar{\Omega}$ has exactly the same expression as before. We obtain
\begin{widetext}
\begin{align}
	|R_{(ii)}|^2 &=\left|
\frac{4\left(B_{12}(B_{12}+B_{22})m_1^2-2B_{11}(B_{12}+B_{22})m_1 m_2
+2B_{11}^2 m_2^2+2B_{11}m_2(B_{12}m_1-B_{11}m_2)\cos(k)\right)\sin^2(k/2)e^{-2ik}}
{\left(B_{12}m_1(e^{ik}-1)+2B_{11}m_2-2B_{11}m_2\cos(k)\right)
\left(B_{12}m_1(e^{ik}-1)-2B_{22}m_1+2B_{11}m_2-2B_{11}m_2\cos(k)\right)}\right|^2\,,\nonumber\\
	|T_{(ii)}|^2 &=\left|
\frac{B_{12}B_{22}m_1^2(1-e^{-2ik})}
{\left(B_{12}m_1(e^{ik}-1)+2B_{11}m_2-2B_{11}m_2\cos(k)\right)
\left(B_{12}m_1(e^{ik}-1)-2B_{22}m_1+2B_{11}m_2-2B_{11}m_2\cos(k)\right)}\right|^2\,.
\label{double}
\end{align}
\end{widetext}

\begin{figure}[th!]
\includegraphics[height=7.5cm]{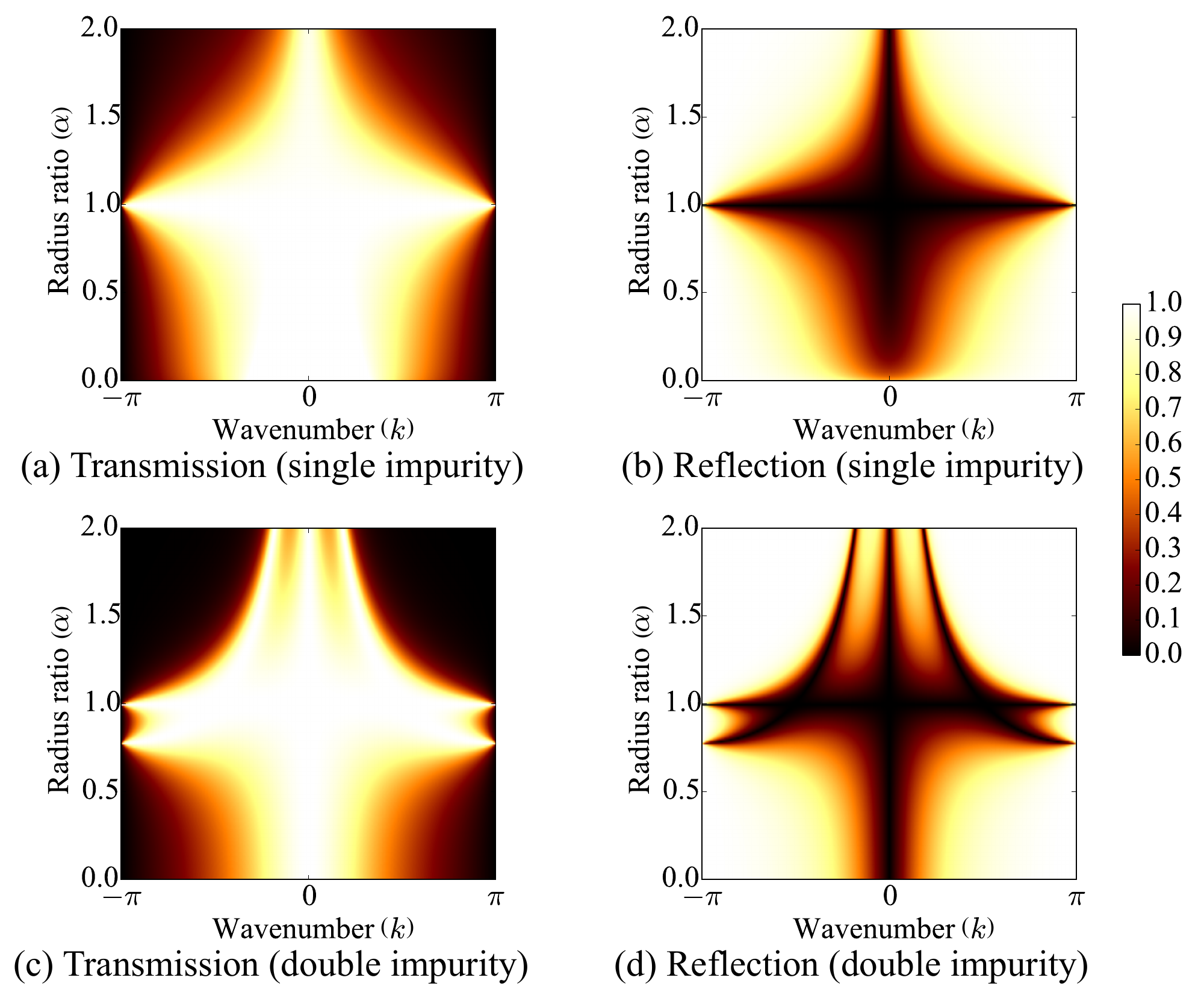}
\caption{(Color online) \textbf{(Left)} Transmission and \textbf{(right)} reflection coefficients for the scattering of a plane wave in a chain with impurities as a function of the wavenumber $k$ 
and the radius ratio $\alpha$. We show examples for \textbf{(top)} a single impurity and \textbf{(bottom)} a double impurity. We describe the physical parameters of the particles in the chain in Table~\ref{tab:property}.
}
\label{f5-t}
\end{figure}

\begin{figure}[t]
\includegraphics[height=7.5cm]{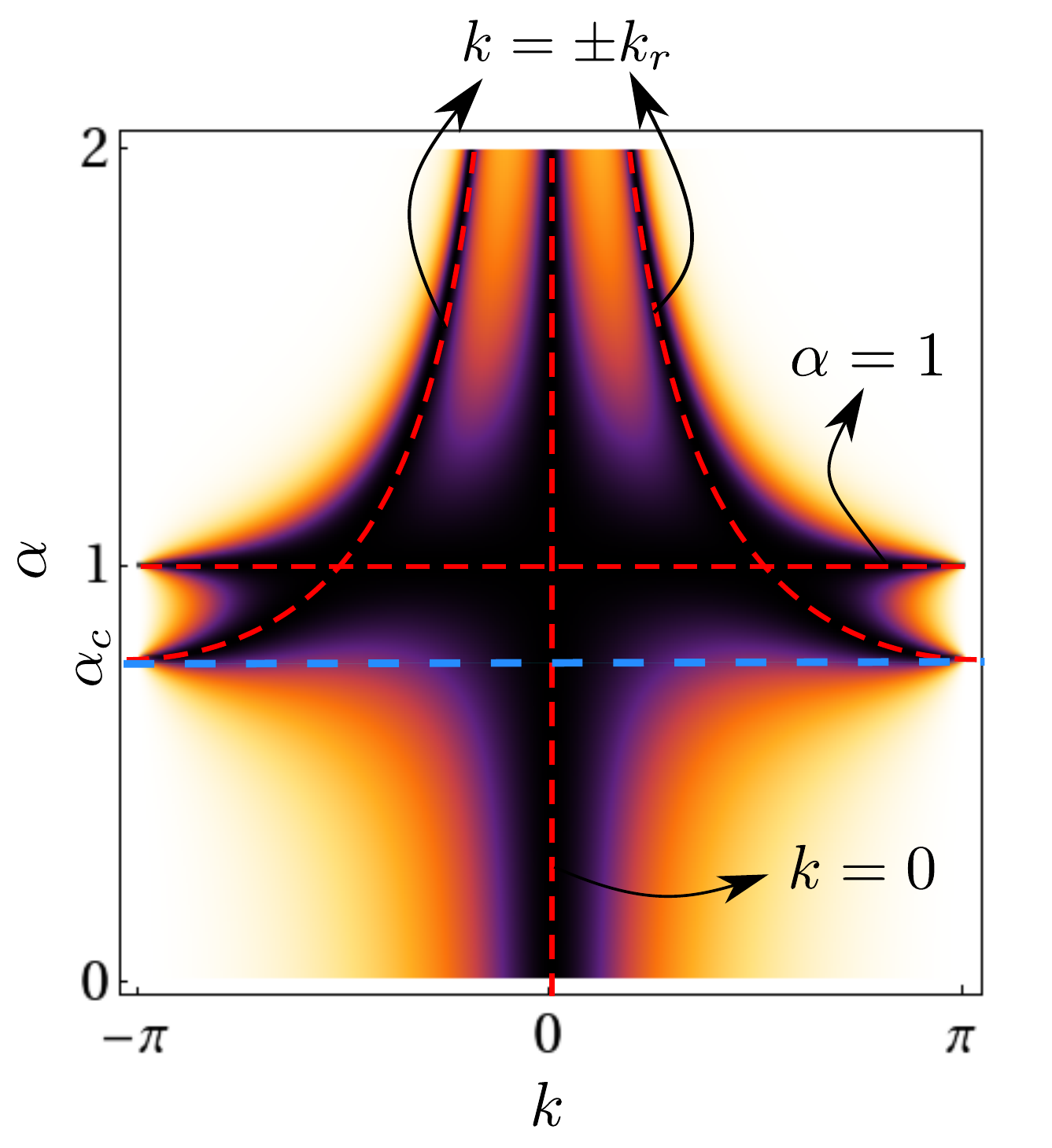}
\caption{(Color online) Reflection coefficient $|R_{(ii)}|^2$ for the double impurity. The red dashed curves indicate the points at which the reflection coefficient is exactly $0$. The resonant wavenumber $k_r$ is given by Eq.~\eqref{reflectionlessmode}.
The blue dashed line highlights the critical value $\alpha_c$; the system has a Ramsauer--Townsend (RT) resonance at wavenumber $k=\pm k_r$ for $\alpha>\alpha_c$.
}
\label{reflectionless}
\end{figure}

In Fig.~\ref{f5-t}, we show the reflection and transmission coefficients as functions of $k$ and $\alpha$. Observe in panels (b) and (d) that there is a black region of {\it reflectionless} modes that can traverse either a single impurity or a double 
impurity almost without modification. For single impurities, the reflection coefficient $|R|^2$ vanishes only when either $\alpha = 1$ or $k=0$. By contrast, for a double impurity, $|R|^2$ vanishes not only when $\alpha = 1$ and $k=0$ but also when $k=\pm k_r\neq 0$ for $\alpha$ larger than some critical value $\alpha_c$. At these resonant values, a wave can be transmitted completely through the impurities (i.e., there is no scattering), and it experiences only a phase shift. Granular crystals thereby yield an analog of the well-known Ramsauer--Townsend (RT) effect~\cite{Sakurai}, which in its traditional form consists of the presence of a sharp minimum in the electron scattering cross-section 
at low energies for scattering with rare gases (such as Xe, Kr, and Ar). Hereafter we use the term ``RT resonance'' to describe the resonance at $k=\pm k_r$. In our case, one can explicitly write $k_r$ in terms of the physical parameters of the system as
\begin{equation}\label{reflectionlessmode}
 	k_{r}= \arccos\left(\phi\right)\,,
\end{equation}
where
\begin{align*}
 	\phi &= \frac{B_{12}B_{22}m_1^2 - 2B_{11}B_{22}m_1 m_2}{2B_{11}m_{2}\left[B_{11}m_2-B_{12}m_1\right]}\\
 &+\frac{B_{11}\left[m_1^2-2m_1m_2+2m_2^2\right]}{2m_{2}\left[B_{11}m_2-B_{12}m_1\right]}\,.
\end{align*}
In Fig.~\ref{reflectionless}, we show $k_r$ and the other relevant values of the reflection coefficient for a double impurity in terms of the parameter $\alpha$. To ensure that $k_r\in \mathbb{R}$, we need $\phi \in \left[-1,1\right]$. In terms of $\alpha$, this implies that the resonant wavenumber $k_r$ exists when $\alpha_c\leq \alpha<\infty$. An interesting feature of $k_r$ is that it can be tuned as a function of $\alpha$ to any value in the interval $[0,\pi]$. In particular, we find that $k_r = \pi$ at $\alpha = \alpha_c$ and $k_r\rightarrow 0$ as $\alpha\rightarrow \infty$. Consequently, one can tune the frequency of the RT resonance to any value in the transmission band $\left[0,\Omega\right]$ of the host granular chain.

In the following subsections, we discuss our computational and experimental results on transmission, and we compare them with our analytical predictions for transmission from Fig.~\ref{f5-t} (obtained using a {\it linear} approximation, as we discussed above).


\subsection{Numerical simulations}\label{sec:simulation}

\begin{figure*}[htbp]
\centerline{ \epsfig{file=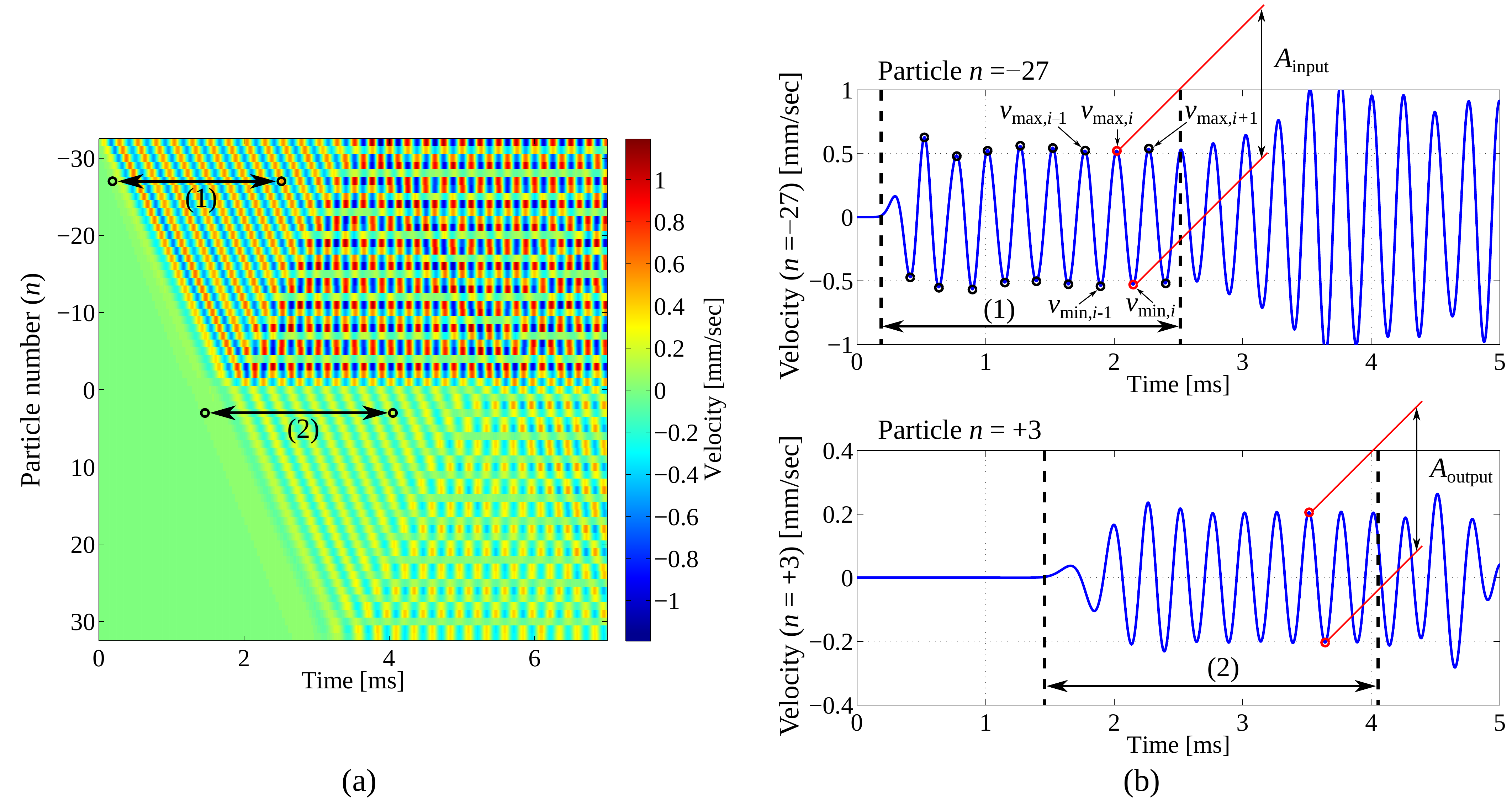,width=0.9 \textwidth} }
\caption{(Color online) \textbf{(a)} Space-time contour plot of particle velocity profiles in a host 63-particle chain in which a double impurity has been inserted between particles $-2$ and $+1$. We use $\alpha = 1.5$ and the parameters in Table~\ref{tab:property} for this numerical simulation. Arrows (1) and (2) indicate the regions that we consider for the calculation of the transmission coefficient. These regions are not affected by the plane waves that reflect from the left or right walls. We also show velocity profiles for particles \textbf{(b)} $n=-27$ and \textbf{(c)} $n=+3$. The dots indicate the maximum and minimum peaks of oscillatory velocity profiles, and the domains of (1) and (2) correspond to the temporal regions marked with (1) and (2) in panel \textbf{(a)}. 
}
\label{fig:Velocity_measurement}
\end{figure*}

For our numerical computations, we solve Eq.~\eqref{Hertz} directly via a Runge--Kutta method (using the {\sc ode45} function in {\sc Matlab}). To quantify the transmission efficiency of the impurity-bearing chains, we analyze velocity profiles of propagating waves under harmonic excitations, as discussed earlier. In Fig.~\ref{fig:Velocity_measurement}(a), we show a space-time contour plot of particle velocities from numerical simulations. In this case, we consider a double impurity (with $\alpha =1.5$) embedded between particles $-2$ and $+1$ [see Fig.~\ref{f4-scattering}(b)] of a 63-particle chain.

The sinusoidal perturbation that we apply to the left end of the chain has a frequency of 4 kHz and an amplitude of 0.35 N. In this scenario, we calculate the magnitude of the particles' maximum displacements to be less than $4.59 \times 10^{-8}$ m. The associated oscillations are two orders-of-magnitude smaller than the static precompression $\Delta_n \approx 1.02 \times 10^{-6}$ from $F_0 = 10$ N, so it is reasonable to assume that the system is operating near the linear regime.

To quantify transmission efficiency, we measure the velocity profiles at specific particles: $n=-27$ for incident waves and $n=+3$ for transmitted waves. We choose these particle locations to allow a sufficiently long spatial interval
between the two nodes in Fig.~\ref{fig:Velocity_measurement}(a). The two-sided arrows (1) and (2) indicate regions
over which the motion is not affected by the presence of
reflections by the chain boundaries. 
In Fig.~\ref{fig:Velocity_measurement}(b), we show the velocity profiles of particles $n=-27$ (top panel) and $n=+3$ (bottom panel). The arrows (1) and (2) again correspond to the temporal domains without interference from wave reflection.

In the temporal plots of velocity profiles, we denote the maxima by $v_{\mathrm{max},i}$ and the minima by $v_{\mathrm{min},i}$, where $i \in \{1, 2, \dots \}$ is the index of the wave peaks in the oscillation. As indicated by the dots in Fig.~\ref{fig:Velocity_measurement}(b), the values of these peaks are not constant even in the designated region before the arrival of the reflected waves. Therefore, we need to extract the steady-state component from the propagating 
plane waves. To do this, we calculate the relative error between a pair of adjacent peaks:
\begin{equation} \label{eq:Error}
	\mathrm{Error}_i=\frac{{{v}_{\max ,i+1}}-{{v}_{\max ,i}}}{{{v}_{\max ,i+1}}}\,.
\end{equation}
We identify the steady-state component of the waves by finding a wave packet with a minimal error. The amplitude $\tilde{A}_i$ of the steady-state velocity component is then
\begin{equation} \label{eq:Error2}
	\tilde{A}_i = {v}_{\max ,i}-{{v}_{\min ,i}}\,.
\end{equation}

By calculating $\tilde{A}_i$ for each peak $i$, we measure the incident wave amplitude $A_{\mathrm{input}}$ and transmitted wave amplitude $A_{\mathrm{output}}$ [see Fig.~\ref{fig:Velocity_measurement}(b)]. Finally, we quantify the transmission coefficient by calculating the ratio of the transmitted wave's velocity amplitude to that of the incident wave:
 \begin{equation} \label{eq:Def_trans}
	\bar{T}_{(i),(ii)} = \frac{{{A}_{\mathrm{output}; \,(i),(ii)}}}{{{A}_{\mathrm{input}; \,(i),(ii)}}}\,,
\end{equation}
where (as mentioned in Sec.~\ref{sec:theory}) the subscripts $(i)$ and $(ii)$, respectively, indicate cases with a single impurity and a double impurity. The transmission coefficient $\bar{T}$, which is written in terms of velocity amplitudes, 
should be equivalent to the displacement ratios introduced in Eq.~\eqref{ansatz} in the ideal situation of harmonic responses of the particles. In the next subsection, we will present our numerical and experimental calculations of $\bar{T}_{(i),(ii)}$.


\subsection{Experimental setup and diagnostics}\label{sec:experiment}

\begin{figure}[htbp]
\centerline{ \epsfig{file=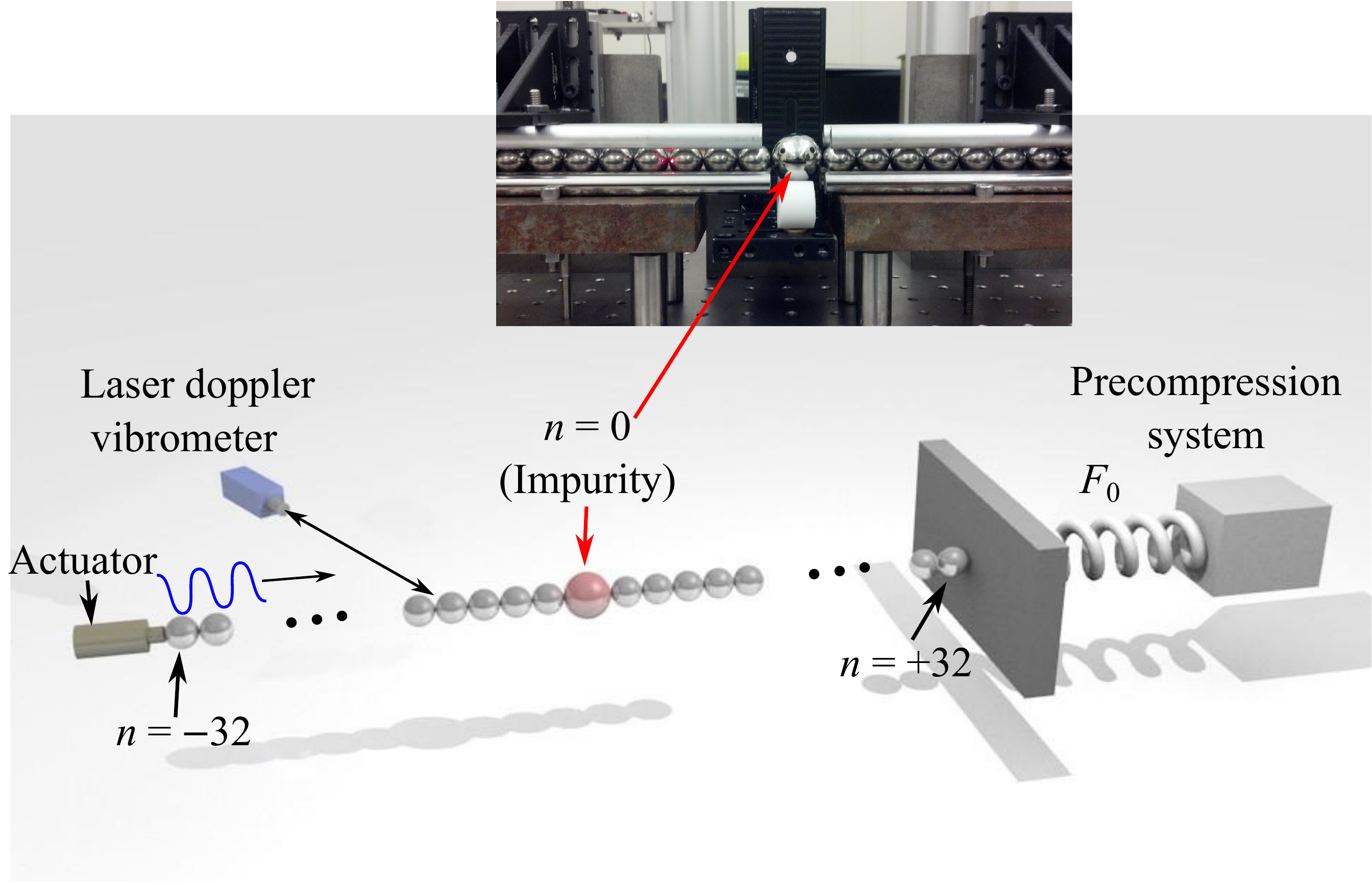,width=0.5 \textwidth} }
\caption{(Color online) Schematic of the experimental setup for a granular chain with a single impurity. In the inset, we show an 
image of the experimental setup.}
\label{fig:Test_setup}
\end{figure}

\begin{figure*}[htbp]
\centerline{ \epsfig{file=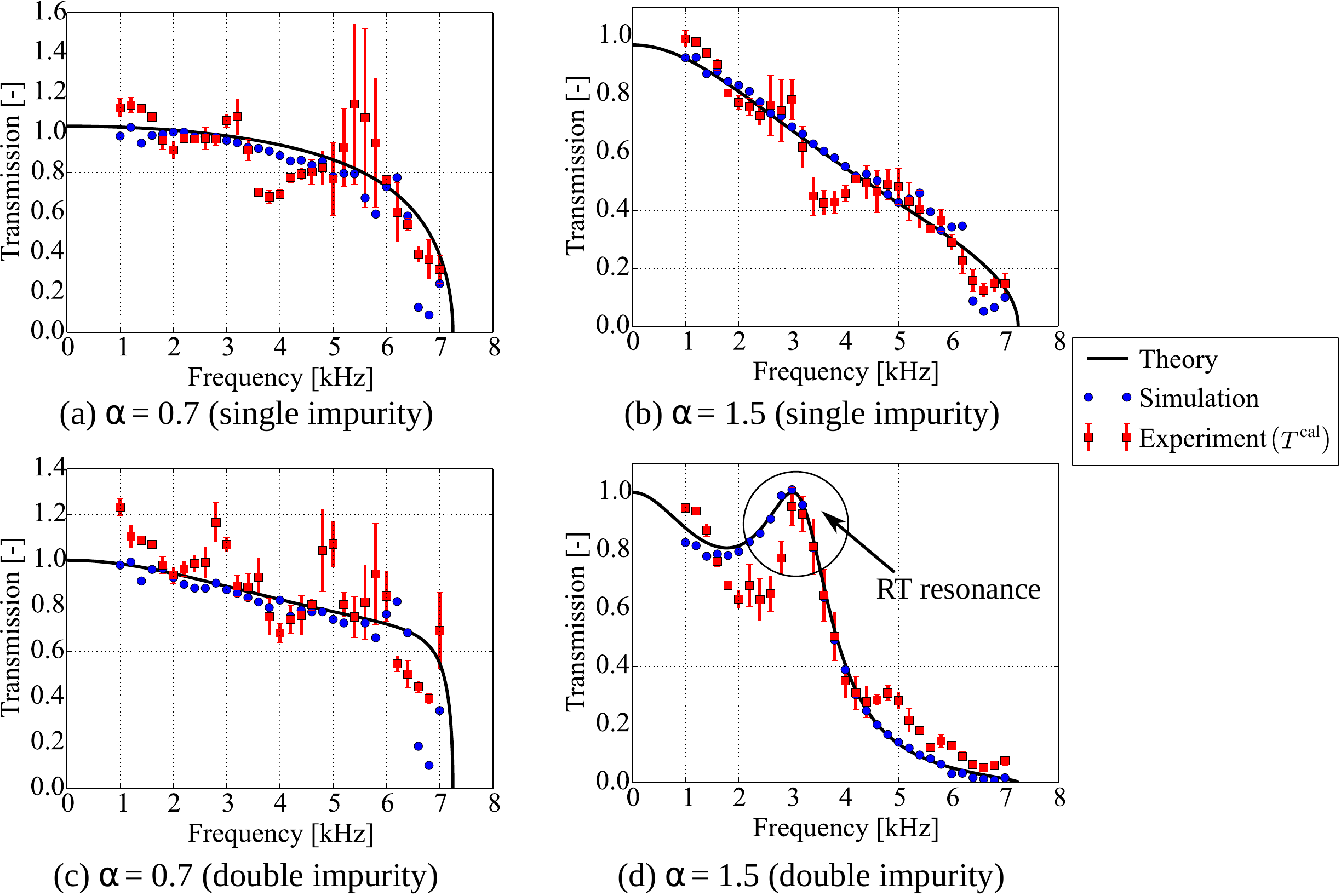,width=0.9 \textwidth} }
\caption{(Color online) Transmission of plane waves in a granular chain with impurities. The radius ratio of the impurity radius to the host particle radius in the host chain is \textbf{(left)} $\alpha=0.7$ and \textbf{(right)} $\alpha=1.5$. We show results for chains with \textbf{(top)} a single impurity and \textbf{(bottom)} a double impurity.
} 
\label{fig:Experiment}
\end{figure*}

We now discuss the results of experiments
in granular chains with a single impurity and a double impurity.  In Fig.~\ref{fig:Test_setup}, we show a schematic to illustrate our experimental setup. We consider a granular chain with 65 spheres: there are 64 type-1 spheres and one impurity in the single-impurity chain, and there are 63 type-1 spheres and 2 impurities in the double-impurity chain. Because of availability limitations, we use distinct materials for type-1 and type-2 particles. However, their material properties are sufficiently similar (see Table~\ref{tab:property}) so that it is permissible to treat them as identical materials. As we show in the inset of Fig.~\ref{fig:Test_setup}, we align the type-1 particles by using four stainless steel rods, and the impurity particle is held in place by an external holder that ensures that its center is aligned with the other particles in the chain. 

To excite the granular chain, we position a piezoelectric actuator on the left side of the chain in direct contact with particle $n = -32$. To generate plane waves in the granular system, we use 
harmonic excitations with a frequency range from 1.0 to 7.0 kHz with a 200 Hz increment. The right end of the chain is compressed by the wall with a static precompression of $F_0$ using a spring 
and linear-stage system. We visualize the propagation of stress waves by measuring the velocity profiles of particles via a non-contact laser Doppler vibrometer (Polytec, OFV-534). See Refs.~\cite{visualization,cho2} for the details of 
this full-field visualization technique.
  
Similar to our numerical approach, we measure the transmission coefficient by estimating the amplitude of the incident ($A_{\mathrm{input}}$) and transmitted ($A_{\mathrm{output}}$) waves. Unlike our numerical simulations, however, the experimental results are susceptible to noticeable attenuation because of dissipation, friction, and slight particle misalignment.  Therefore, we calibrate our experimental results by normalizing them with respect to the measurement results obtained from a homogeneous particle chain. The calibrated transmission coefficient is thus
\begin{equation} \label{eq:Def_cal_trans}
	{\bar{T}^{\mathrm{cal}}_{(i),(ii)}}
	=\frac{{\bar{T}_{(i),(ii)}}}{{\bar{T}_{\alpha =1}}}
	=\frac{A_{\mathrm{output}; \,(i),(ii)}}{A_{\mathrm{output}; \,\alpha =1}}\,,
\end{equation}
where $\bar{T}_{(i),(ii)}$ is the transmission coefficient for single-impurity and double-impurity chains based on Eq.~\eqref{eq:Def_trans}, and $\bar{T}_{\alpha =1}$ is the transmission coefficient for a homogeneous chain (i.e., for $\alpha =1$).

\begin{table}[htb]
  \begin{center}
    \caption{Properties of type-1 and type-2 particles.} 
    \begin{tabular}{l|r|r} \hline
      & Type-1 & Type-2 (impurity) \\ \hline \hline
      Material & 440C & AISI 52100 \\
      Elastic modulus & $E_1=204$ GPa & $E_2=210$ GPa \\
      Poisson ratio & $\nu_1 = $ 0.28 & $\nu_2 = $0.30 \\
      Density & $\rho_1=7.80$ g/cm$^3$ & $\rho_2=7.81$ g/cm$^3$ \\
      Radius & $r_1=9.525$ mm & $r_2=\alpha r_1$ \\ \hline
    \end{tabular}
    \label{tab:property}
  \end{center}
\end{table}


\section{Comparison between analytical, numerical, and experimental results}\label{sec:comparison} 

We now compare our analytical results with numerical simulations and experimental data for the radius ratios $\alpha=0.7$ and $\alpha=1.5$. In Figs.~\ref{fig:Experiment}(a,b), we show our results for the transmission coefficients for a single-impurity chain. In Figs.~\ref{fig:Experiment}(c,d), we present our results for a double-impurity chain. In these plots, black solid curves indicate the analytical predictions from Eqs.~\eqref{single} and \eqref{double}, blue dots indicate the results of simulations obtained by solving Eq.~\eqref{Hertz}, and red squares give the experimental results after calibration using Eq.~\eqref{eq:Def_cal_trans}. 

For a single-impurity chain, the transmission coefficient has a decreasing trend as we increase the excitation frequency.  This supports our prediction from Fig.~\ref{f5-t}(a). The slope of the decrease depends on the mass ratio. When $\alpha = 0.7$, the decreasing trend starts slow, but there is a rapid drop around the cutoff frequency of 7.25 kHz that we obtained analytically from the formula $\Omega = \sqrt{\frac{4B}{m}}$. For $\alpha = 1.5$, the decrease has a near-linear trend throughout the frequency pass band. In Figs.~\ref{fig:Experiment}(a,b), we observe these trends in both numerics and experiments. However, as we will discuss shortly, there are some differences in the experiments as compared to the simulations and theoretical predictions.
 
For a double-impurity chain, we obtain more interesting, 
potentially non-monotonic behavior. When $\alpha = 0.7$, we observe, broadly speaking, a decrease of transmission efficiency as the frequency increases; this is reminiscent of the single-impurity chain. However, for the mass ratio $\alpha = 1.5$, the transmission coefficient has a pronounced double-peak shape in the frequency pass band. In particular, our analytical results for transmission predict a resonant mode at an excitation frequency of about 3.0 kHz. This leads to complete transmission of plane waves despite the existence of impeding double impurities. This ``cloaking'' mode is notable, and we observe it in both experiments and numerical simulations [see Figs.~\ref{fig:Experiment}(c,d)]. However, we again note
that quantitative differences exist despite the accurate qualitative description of the experiment and
the numerical corroboration.


As we have just discussed, our analytical predictions match reasonably well with the results of our numerical simulations and experimental findings, especially for frequencies between 1.0 and 4.0 kHz. By comparing analytical predictions and experimental results around 3.0 and 4.0 kHz, however, we observe some discrepancies that are 
not noticeable when comparing analytical and numerical calculations. They probably stem from experimental errors, such as a potential slight misalignment of the external holder and, perhaps more notably, 
an intrinsic frequency response of a 
piezo actuator. 

For higher frequencies, especially between 6.0 and 7.0 kHz, we observe an especially noticeable discrepancy when comparing the theoretical predictions to the numerical and experimental results. 
[For example, see Figs.~\ref{fig:Experiment}(b,c).] We believe that this arises due to transient waveforms --- and specifically 
due to wave localization --- in the vicinity of the excitation particle (i.e., at the left end of the chain). 
If one excites a granular chain from a stationary state, the propagating waves include a wide range of frequencies near the excitation frequency. If the excitation frequency is close to the cutoff frequency, then incident waves whose frequencies are larger than the cutoff frequency will not propagate but will instead be localized at the excitation particle in the form of evanescent waves. Such perturbations result in transient behavior in the form of propagating waves, often in modulated waveforms in the time domain. This, in turn, affects the calculation of transmission coefficients in numerical simulations and experiments.  In both cases, we examine the dynamics in subsets of the chains for small propagation times to avoid the effects of reflection from the right boundary. See the Appendix for further details.


\section{Multiple impurities}\label{sec:multiple}

An interesting application of the RT resonance, which we discussed in Sec.~\ref{sec:theory} for scattering with a double impurity, is its extension to systems with multiple double impurities. In particular, it is interesting to examine systems in which multiple impurities are either periodically or randomly distributed within a host homogeneous chain. 
A fascinating question arises: can reflectionless modes still occur?

When considering multiple impurities in a host granular chain, the formalism of transfer matrices provides a useful framework to study transmission of waves through the entire system~\cite{randombook}. Following recent work by Zakeri et al.~\cite{LepriLevy}, we assume stationary plane waves $u_n(t) = w_n e^{i\omega t}$ as in Sec.~\ref{sec:theory}. Equation~\eqref{Linear} then leads to
\begin{equation}\label{transfermatrixeq}
	w_{n+1} = \frac{\left[(B_n+B_{n+1})-m_n\omega^2\right]}{B_{n+1}} w_n - \frac{B_{n}}{B_{n+1}}w_{n-1}\,,
\end{equation}
which generates the modes given a seed $\{w_{-N},w_{-N+1}\}$. For $\omega = 0$, Eq.~\eqref{transfermatrixeq} reduces to 
\begin{equation}
	w_{n+1} = \frac{B_n}{B_{n+1}}\left(w_n-w_{n-1}\right) + w_n\,.
\end{equation}
Thus, for any distribution of particles in the chain, the seed $w_{-N}=w_{-N+1}$ implies that $w_n=w_{-N}$ for all $n$. This explains why the reflection coefficients are exactly $0$ at $\omega =k=0$ for both single and double impurities (see Fig.~\ref{f5-t}).


\begin{figure}[t]
\includegraphics[height=2.5cm]{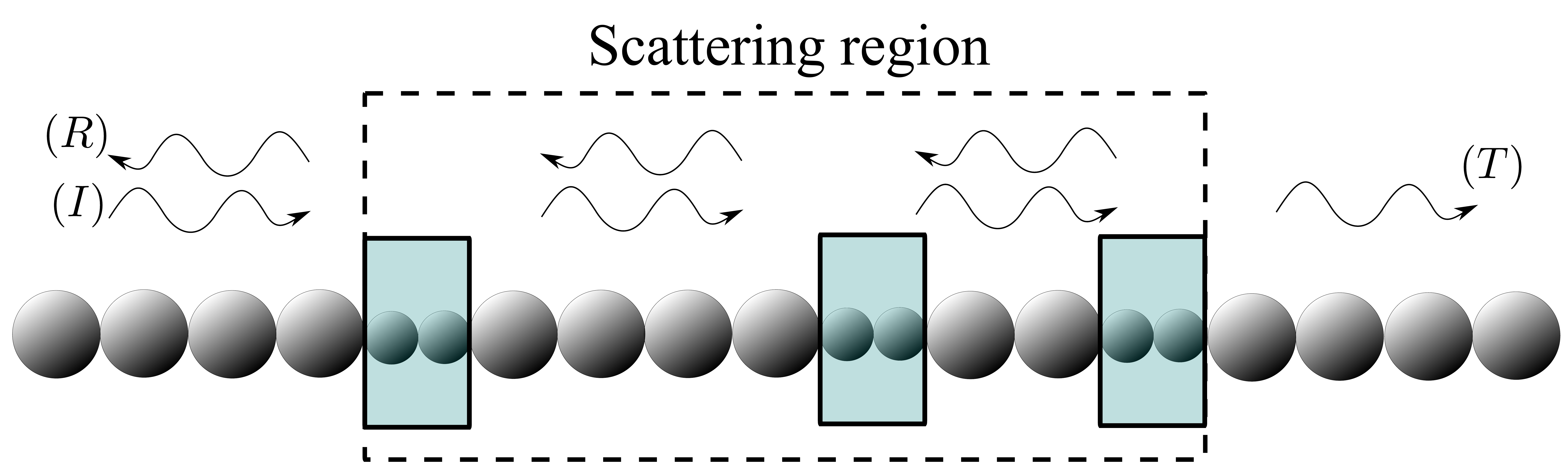}
\caption{(Color online) Schematic of a host homogeneous granular chain with multiple double impurities. The incident wave is $(I)$, the reflected wave is $(R)$, and the transmitted wave is $(T)$. We highlight impurities in solid turquoise boxes, and we indicate the scattering region with the dashed box.
}
\label{multiple}
\end{figure}


In the absence of impurities, Eq.~\eqref{transfermatrixeq} generates propagating waves for any $\omega\in[0,\Omega]$; this is, at least
true in the infinite domain, while for a finite domain, only the 
wavenumbers conforming to the specific boundary conditions, and
the associated frequencies get selected. Once we add impurities, the iterative process to generate
such propagating waves is the same until we reach what we call a ``scattering region'' (see Fig.~\ref{multiple}). In this region, multiple scatterings occur because the presence 
of impurities has broken discrete translation symmetry, and successive interferences can then lead to complicated dynamics that depend on the distribution of impurities. A 
particular example of this phenomenon was investigated recently in the context of disordered granular chains~\cite{martinez2014}. When the distribution of impurities is such that impurities are well separated from each other,
one can reformulate the transmission problem [given by Eq.~\eqref{transfermatrixeq}] through the entire scattering region as a sequence of transfer problems from each segment of
a granular chain through an impurity to the next segment. Thus, an incident plane wave $w_n = I e^{ik_r n}$ with wavenumber $k=k_r$ and amplitude $I$ transforms into $T^{(1)}e^{ik_r n}$ after a scattering event because no 
reflected waves are generated during the scattering at $k=k_r$. By considering each impurity, we obtain the sequence 
$Ie^{ik_r n}\rightarrow T^{(1)}e^{ik_r n}\rightarrow T^{(2)}e^{ik_r n}\rightarrow \cdots \rightarrow T^{(L)} e^{ik_r n}$, where arrows denote the transmission of 
the wave through the impurities and $T^{(j)}$, with $j \in \{1,2,\ldots,L$\}, denotes the transmitted-wave amplitudes, 
which are are given by Eq.~\eqref{double}. Consequently, reflectionless modes can be supported by the chain in the form
\begin{equation}
	w_n = \left\{
		\begin{array}{lcc}
		I e^{ik_r n}\,,&\quad\text{if}&n\leq n_1\\
		T^{(1)}e^{ik_r n}\,,&\quad\text{if}& n_1 < n \leq n_2\\
		\vdots& &\\
		T^{(L)}e^{ik_r n}\,,&\quad\text{if}& n_L < n
		\end{array}
	\right.\,,
\label{ansatz2}
\end{equation}
where $n_j$ (with $j \in \{1,2,\ldots,L\}$) represents the positions of the $j$th impurity in the host homogeneous chain.
For other modes, transmission through the scattering region depends on the frequency $\omega$. Based on our analysis in Sec.~\ref{Sec2}, we expect that 
transmission decays rapidly as one approaches the upper band edge $\Omega$. By contrast, plane waves slowly attenuate for frequencies near $0$.


\begin{figure}[t]
\includegraphics[height=8cm]{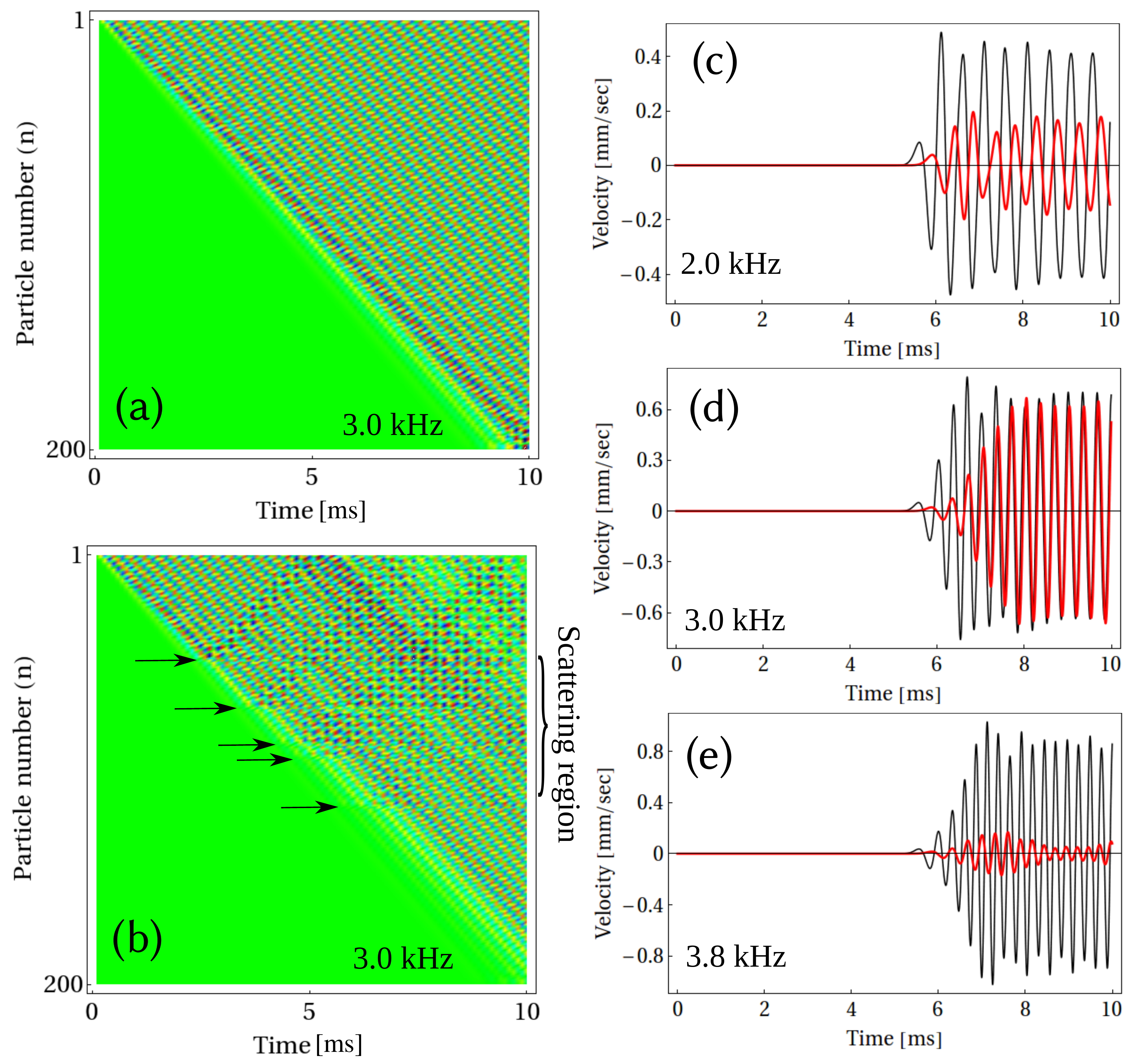}
\caption{{Color online} (a) Space-time contour plot of the normalized velocity for a homogeneous chain with $N=200$ particles and an excitation frequency of $3.0$ kHz. 
(b) The same plot as in panel (a), but with five double impurities located at positions $n=50$, $n = 72$, $n = 90$, $n = 96$, and $n = 118$. The radius ratio is $\alpha = 1.5$, and arrows indicate the position of the impurities.  Panels (c)--(e) show the velocity at particle $n=130$ for excitation frequencies of (c) $2.0$ kHz, (d) $3.0$ kHz, and (e) $3.8$ kHz. The black (large-amplitude) curves are associated with the homogeneous chain [panel (a)], and the red (smaller-amplitude) curves are associated with the chain with the impurities [panel (b)].
}
\label{multiple2}
\end{figure}

To corroborate that these effects arise in strongly precompressed granular crystals, we numerically integrate Eq.~\eqref{Hertz} using the same parameters as in 
Sec.~\ref{sec:simulation}, but this time we randomly place double impurities within a scattering region around the middle of a chain. Specifically, we set the radii of
each pair of consecutive particles within the scattering region to be $r_1$ with probability $1/2$ and $r_2$ with probability $1/2$. We consider a chain with $N=200$ particles, and we define the scattering region to be between particles $n=50$ and $n=120$. We set the radius ratio to be $\alpha=1.5$, and we generate $10^3$ random chains.
One can calculate the frequency of the RT resonance using Eq.~\eqref{reflectionlessmode}, which in this case gives $f_r \approx 3.0$ kHz. As we predicted, when the system is driven at this frequency, waves experience a phase shift due to the scattering, but the amplitude is consistently transmitted almost without 
modification through the scattering region of the random chains. 
However, when we move away from the frequency of the RT resonance, the transmission decays. 
We observe this directly by measuring the velocity of a particle right after the scattering region (see Fig.~\ref{multiple2}). We compare the temporal evolution of the velocities of the particles for a homogeneous chain and a chain with 
five double impurities. In panels (c) and (e), we observe attenuation in the magnitude of the velocity due to the presence of impurities in the chain. In panel (d), when the system is driven at $3.0$ kHz, the wave is clearly delayed in the perturbed chain compared with the homogeneous one, although the magnitude of the velocity is about 
the same for both chains. As predicted, we observe the RT resonance even in granular chains with multiple double impurities.


\section{Conclusions}\label{conclusion}

In the present work, we examined the scattering of waves by single impurities and double impurities in granular chains. We started by exploring the linear scattering problem motivated by the context of strongly precompressed granular chains with either a single impurity or a double impurity. We derived analytical
formulas to show that the scattering is markedly different for the different impurity configurations. 
For single-impurity chains, we showed that the transmission coefficient $|T|^2$ decays monotonically with $k$ (and hence with the frequency $\omega$). We also found that $|T|^2\rightarrow 0$ as one approaches the 
band-edge frequency of the host homogeneous chain. By contrast, for a double-impurity chain, we showed that an effect analogous to the Ramsauer--Townsend resonance takes place at $k=k_r\in [0,\pi]$ and for a specific region of parameter space. We demonstrated that one can tune the frequency of this RT resonance to any value within the transmission band of the host homogeneous chain. 

We compared our analytical results to numerical computations and laboratory experiments, and we obtained good agreement. In our experiments, we used non-contact laser Doppler vibrometry to obtain a full-field visualization of plane waves propagating in a granular chain. This allowed us to observe the RT resonance for double impurities
in a granular chain by directly measuring the transmission coefficient associated with the scattering. We also discussed how this RT resonance can be responsible for the emergence of reflectionless modes in systems with multiple (either ordered or disordered) double impurities. We demonstrated this reflectionless transmission using numerical simulations.

Our study paves the way for a systematic study of the properties of Ramsauer--Townsend resonances in granular crystals. One could study such resonances in the context of more impurities, or more systematically in the case
of (ordered or disordered) distributions of impurities in such granular media. One possible application of RT resonances in granular crystals is embedding foreign objects, such as sensors, in systems so that they induce minimal interference with the existing structures. 
It is also of considerable interest to explore disordered granular crystals, rather than merely placing a disordered segment in otherwise homogeneous chains. In 1D disordered granular crystals, the recent numerical predictions of superdiffusive transport and other features~\cite{martinez2014,theocharis2015} are especially interesting to further explore. Such efforts are currently in progress.


\section*{Acknowledgements}

A.J.M. acknowledges partial support from CONICYT (BCH72130485/2013). J.Y. and H.Y. acknowledge the support of ONR (N000141410388), NSF (CMMI-1414748 and -1553202), and ADD of Korea (UD140059JD). P.G.K gratefully acknowledges support from US-AFOSR under grant FA9550-12-1-0332, and the ERC under FP7, Marie Curie Actions, People, International Research Staff Exchange Scheme (IRSES-605096). J.Y. and P.G.K. also acknowledge support from US-ARO under grant (W911NF-15-1-0604).




\appendix
\section*{Appendix: Experimental verifications of wave localization}

\begin{figure*}[htbp]
\centerline{ \epsfig{file=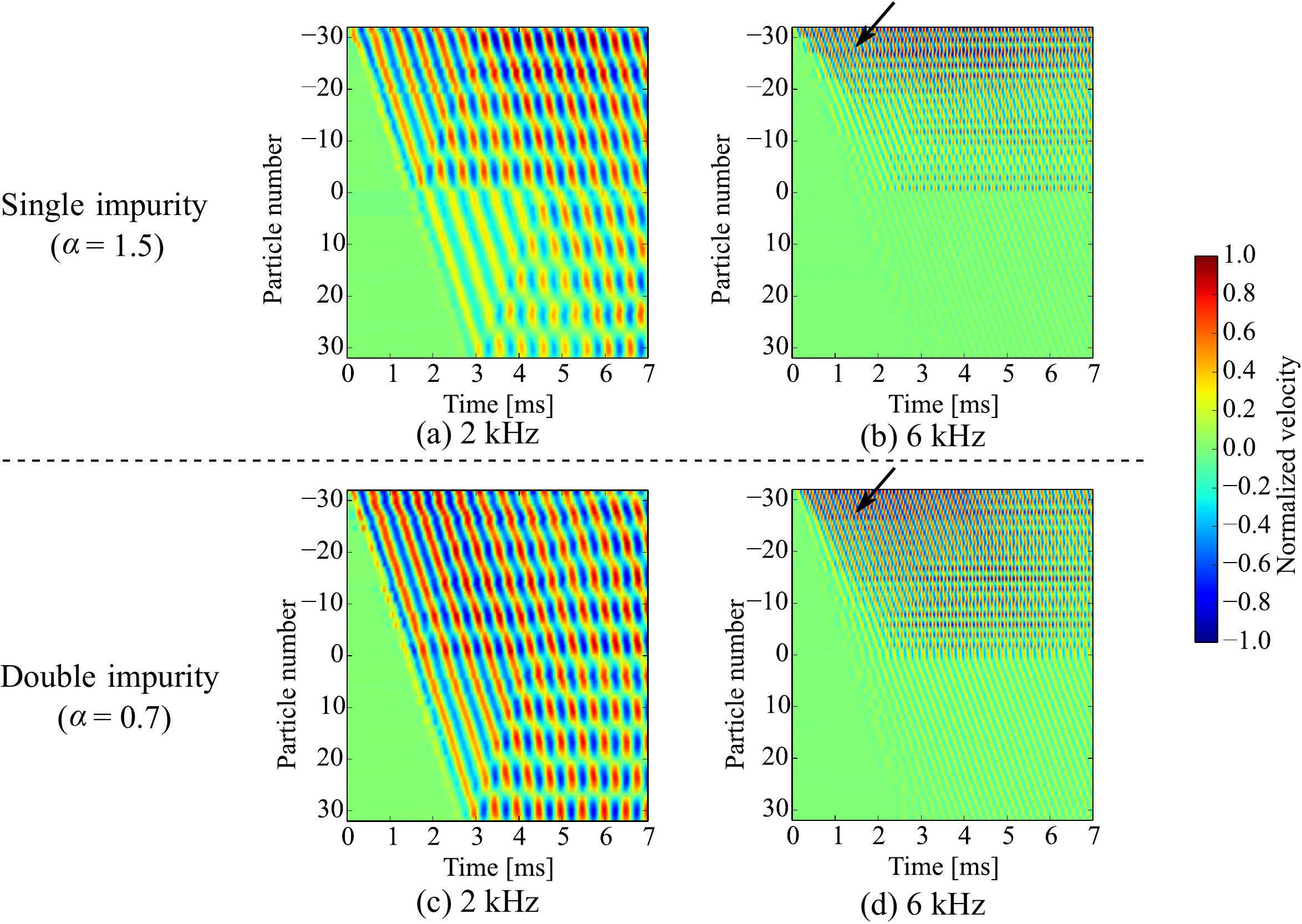,width=0.9 \textwidth}}
\caption{(Color online) Space-time contour plot of normalized velocity for experiments with \textbf{(top)} a single-impurity chain with $\alpha=1.5$ and \textbf{(bottom)} a double-impurity chain with $\alpha=0.7$. In each case, we normalize the measured velocities with respect to the maximum velocity. We use excitation frequencies of \textbf{(left)} 2 kHz and \textbf{(right)} 6 kHz. The arrows point to incidents of wave localization.
}
\label{fig:perturbation}
\end{figure*}

To examine the wave localization that we mentioned in Sec.~\ref{sec:comparison}, we perform experiments to visualize full-field velocity profiles of all particles in a chain. In Fig.~\ref{fig:perturbation}, we show space-time contour plots of velocity profiles for (top row) the single-impurity chain with $\alpha=1.5$ and (bottom row) the double-impurity chain with $\alpha=0.7$. 
 We use excitation frequencies of (left panel) 2.0 kHz and (right panel) 6.0 kHz. The experimental results in Fig.~\ref{fig:perturbation} require measurements of the motion of individual particles followed by synchronization of all measured data, because the laser Doppler vibrometer scopes only a single particle's motion at a time. In each case, after we collect all data, we normalize the measured values of particles' velocities with respect to the maximum velocity component.

As we indicate with the arrows in the right panels of Fig.~\ref{fig:perturbation}, we observe localization in our single-impurity and double-impurity experiments when the excitation 
frequency is 6.0 kHz. We do not find such a distinctive localization for the 2.0 kHz excitation [see Fig.~\ref{fig:perturbation}(a,c)]. Again, as explained in Sec.~\ref{sec:comparison}, this is 
due to the inevitable perturbation of ``beyond-cutoff frequency" components of stress waves in experiments when we excite the system near the cutoff frequency. Incident waves whose frequencies are close to the cutoff frequency cause this perturbation even when the precompression $F_0$ is large enough or the excitation amplitude is small enough to 
 remain near the linear regime of the granular chain. This wave localization 
contributes to the discrepancy between the experimental and analytical data around 6.0 kHz in 
 Fig.~\ref{fig:Experiment}. In our numerical simulations, we also observe wave localization at the edge of the chain, and we thereby obtain a dip in our transmission data around 6.0 kHz that systematically appears for different values of $\alpha$ and different chain lengths.

\end{document}